\documentclass[prl,twocolumn,showpacs,superscriptaddress]{revtex4}

\usepackage{amsmath}
\usepackage{graphicx}
\usepackage{amssymb}

\begin{document}

\newcommand{\ie}{i.e.\ }
\newcommand{\ket}{\ensuremath{\rangle}}
\newcommand{\bra}{\ensuremath{\langle}}
\newcommand{\tp}{\ensuremath{\otimes}}
\newcommand{\cnot}{\textsc{cnot}}
\newcommand{\swap}{\textsc{swap}}
\newcommand{\m}{\ensuremath{\mathbf{m}}}
\newcommand{\n}{\ensuremath{\mathbf{n}}}
\newcommand{\z}{\ensuremath{\mathbf{z}}}
\newcommand{\maxi}{\ensuremath{\mathrm{max}}}
\renewcommand{\l}{\left}
\renewcommand{\r}{\right}
\renewcommand{\[}{\left[}
\renewcommand{\]}{\right]}
\renewcommand{\(}{\left(}
\renewcommand{\)}{\right)}
\newcommand{\eq}{\equiv}

\title{A practical scheme for quantum computation with any two-qubit
entangling gate}

\begin{abstract}
Which gates are universal for quantum computation? Although it is
well known that certain gates on two-level quantum systems
(\emph{qubits}), such as the controlled-{\sc not} ({\sc cnot}),
are universal when assisted by arbitrary one-qubit gates, it has
only recently become clear precisely what class of two-qubit gates
is universal in this sense.  Here we present an elementary proof
that \emph{any} entangling two-qubit gate is universal for quantum
computation, when assisted by one-qubit gates.  A proof of this
important result for systems of \emph{arbitrary} finite dimension
has been provided by J.~L.\ and R.~Brylinski
[arXiv:quant-ph/0108062, 2001]; however, their proof relies upon a
long argument using advanced mathematics.  In contrast, our proof
provides a simple \emph{constructive} procedure which is close to
\emph{optimal} and experimentally practical [C.\ M.\ Dawson and
A.\ Gilchrist, online implementation of the procedure described
herein (2002), \texttt{http://www.physics.uq.edu.au/gqc/}].
\end{abstract}


\author{Michael~J.~Bremner}
\author{Christopher~M.~Dawson}
\author{Jennifer~L.~Dodd}
\author{Alexei~Gilchrist}
\affiliation{Centre for Quantum Computer Technology and Department of
Physics, The University of Queensland, QLD 4072, Australia}
\author{Aram~W.~Harrow}
\affiliation{Centre for Quantum Computer Technology and Department of
Physics, The University of Queensland, QLD 4072, Australia}
\affiliation{MIT Physics, 77 Massachusetts Ave., Cambridge MA 02139
USA}
\author{Duncan~Mortimer}
\author{Michael~A.~Nielsen}
\author{Tobias~J.~Osborne}
\affiliation{Centre for Quantum Computer Technology and Department of
Physics, The University of Queensland, QLD 4072, Australia}
\maketitle


A great deal of work has been done to determine what physical
resources are capable of universal quantum computation. It is well
known that certain gates on two-level quantum systems such as the
controlled-{\sc not}, are universal when assisted by arbitrary
one-qubit gates~\cite{Barenco95}. It has also been shown that
\emph{almost any} gate on two $d$-level quantum systems
(\emph{qudits}), together with its swapped version, is universal
for quantum computation without the aid of one-qudit gates
\cite{Lloyd95,Deutsch95}.  However, this result does not
explicitly specify which two-qudit gates are universal, requires
the ability to apply the given gate in two different ways, and the
resulting procedure is not practical, requiring large numbers of
gates.

Recently, several authors have considered conditions for
universality when only a single {\em fixed} multi-qudit
interaction, together with one-qudit gates, is allowed
\cite{Dodd02,Dur01,Wocjan02,Bennett01,Vidal01,Nielsen01b}. They
have shown that any interaction that can create entanglement
between any pair of qudits is universal for quantum computation.
Theoretical schemes for quantum computation based on this have
been found, but they are not of practical utility.  In order to
make the simulations exact the given interaction is modified by
one-qubit gates which must be applied so that the period of
evolution between them is infinitesimal.  To simulate evolution
for some non-infinitesimal time $t$ the error is controlled by
concatenating a large number $n$ of periods of evolution for a
small time $t/n$.  Although some such schemes minimize the amount
of time required to do the simulation \cite{Bennett01,Vidal01},
the required number of one-qubit gates is enormous; an optimistic
example of simulating a \cnot\ to accuracy only $10^{-3}$ requires
approximately $10^4$ one-qubit gates \cite{Dodd02}.  Thus, the
one-qubit gates must be performed with rapidity and accuracy which
are vastly more demanding than the standard requirements for
quantum computation.  We note, however, that Hammerer, Vidal, and
Cirac~\cite{Hammerer02} have obtained a practical scheme for a
restricted class of symmetric Hamiltonians with no self-energy.

In contrast, the model we consider does not allow an interaction
to be interrupted by one-qubit gates at arbitrary times.  We allow
only a fixed entangling two-qubit \emph{gate} $U$ and arbitrary
one-qubit gates between applications of $U$.  Our proof that any
such gate is universal provides an explicit method \cite{Dawson02}
for implementing a \cnot\ exactly, using a small number of
one-qubit gates which is \emph{fixed} for any given gate $U$.  For
the optimistic example mentioned above, this method requires only
approximately $10$ one-qubit gates.  This represents a saving of a
factor of $10^3$; typical savings will be much greater.  The only
limit to the accuracy achieved in practice is due to the accuracy
with which the required one-qubit gates are calculated.  This
limit is inherent in any procedure for computation, but because of
the constant number of one-qubit gates required by our scheme, the
induced errors will depend only in a constant way on these
inaccuracies.  Combined with the fact that the number of uses of
$U$ is near-optimal, this suggests that our scheme will be of
practical utility.

We begin the description of our construction with some convenient
definitions:
\begin{itemize}
\item We say that a gate is \emph{universal} if it can be used to
perform universal quantum computation on two qubits when assisted
by arbitrary one-qubit gates. \item Suppose $U=(A_1\tp
B_1)V(A_2\tp B_2)$.  Since we have the ability to do arbitrary
one-qubit gates, being able to perform $U$ allows us to perform
$V$, and vice versa.  Whenever this is the case, we say that $U$
and $V$ are \emph{equivalent} and write $U\equiv V$. \item A gate
$U$ is \emph{entangling} if it can create entanglement between two
systems initially in a product state. \item
Following~\cite{Brylinski01}, we define $U$ to be \emph{primitive}
if $U$ is a product of one-qubit gates or if $U$ is equivalent to
the gate interchanging the two qubits (\swap); otherwise $U$ is
\emph{imprimitive}.  We will see that, for two qubits, the class
of imprimitive gates is exactly the class of entangling gates.
\end{itemize}

We now prove the qubit case of the result in~\cite{Brylinski01}:
\begin{quote}
A two-qubit gate $U$ is universal if and only if it is imprimitive,
or, equivalently, if and only if it is entangling.
\end{quote}
\textbf{Proof:} A brief summary of our proof is as follows: We use
two non-trivial facts.  The first is that \cnot\ is universal
\cite{Barenco95}.  The second is the \emph{canonical
decomposition} \cite{Khaneja01, Kraus00} for any two-qubit gate
$U$:
\begin{equation} \label{eq:canonical}
U=(A_1\tp B_1)e^{i(\theta_xX\tp X+\theta_yY\tp Y+\theta_zZ\tp
Z)}(A_2\tp B_2),
\end{equation}
where $X,Y,Z$ are the Pauli sigma matrices, $A_j,B_j$ are
one-qubit gates, and
$-\frac{\pi}{4}<\theta_{\alpha}\le\frac{\pi}{4}$
(see~\cite{Kraus00} for a simple proof).  Both of these facts have
proofs which are somewhat detailed but elementary and
constructive.  Our strategy is to show that any imprimitive gate
$U$, together with one-qubit gates, can be used to implement
$W=e^{i\phi Z\tp Z}$ where $0<|\phi|<\frac{\pi}{2}$.  We then show
that $W$ can be used, together with one-qubit gates, to exactly
implement \cnot, which proves that $W$, and therefore $U$, is
universal.  Finally, since any universal gate is entangling, and
any entangling gate is imprimitive, it follows that the class of
entangling gates is exactly the class of imprimitive gates.

We define $V=e^{i(\theta_xX\tp X+\theta_yY\tp Y+\theta_zZ\tp Z)}\equiv
U$.  First, note that primitive gates have either
$\theta_x=\theta_y=\theta_z=0$ (corresponding to $U$ being a product
of one-qubit gates), or $\theta_x=\theta_y=\theta_z=\frac{\pi}{4}$
(corresponding to $U\equiv\swap$), so we need not consider these
cases.

Suppose $U$ is imprimitive, in which case at least one of the
$\theta_\alpha$ is non-zero.  We will show that in all cases $V$, and
hence $U$, may be used with one-qubit gates to implement a \cnot\ and
is therefore universal.  In each case, we use $V$ to obtain a gate of
the form $W=e^{i\phi Z\tp Z}$, $0<|\phi|<\frac{\pi}{2}$.  Note that we
may assume $|\theta_z|\ge|\theta_x|\ge|\theta_y|$ since the
$\theta_\alpha$ may be relabeled by conjugating $V$ by the primitive
gates $H\tp H$ and $S\tp S$ where
$H=\frac{1}{\sqrt2}\[\begin{smallmatrix}1&1\\1&-1\end{smallmatrix}\]$
and $S=\[\begin{smallmatrix}1&0\\0&i\end{smallmatrix}\]$.

First, consider the two special cases where either one or two of
$\theta_x,\theta_y,\theta_z$ are $\frac{\pi}{4}$ and the remainder are
0.  Suppose that $\theta_z=\frac{\pi}{4}$ and $\theta_y=\theta_x=0$.
Then $V=e^{i\frac{\pi}{4}Z\tp Z}$ and is hence already of the required
form.  For the second special case, $\theta_z=\theta_x=\frac{\pi}{4}$
and $\theta_y=0$.  Noting that $V^8=I$, and thus $V^7=V^\dag$, we use
the one-qubit gate $e^{i\frac{\pi}{4}X\tp I}$ to obtain
$Ve^{i\frac{\pi}{4}X\tp I}V^7=e^{i\frac{\pi}{4}V X\tp I
V^\dag}=e^{i\frac{\pi}{4}Y\tp Z}\equiv e^{i\frac{\pi}{4}Z\tp Z}$,
which is of the required form.

Secondly, consider the more general case, $\theta_z\neq\frac{\pi}{4}$.
Now
\begin{equation}
(I\tp Z)V(I\tp Z)V=e^{2i\theta_zZ\tp Z}=e^{i\phi Z\tp Z}=W
\end{equation}
where $0<|\phi|<\frac{\pi}{2}$, as required.

Simple algebra shows that $W$ is equivalent to a controlled rotation
about the $z$-axis:
\begin{eqnarray} \label{eq:controlled_rot}
e^{i\phi Z\tp Z}&=&|0\ket\bra0|\tp e^{i\phi Z}+|1\ket\bra1|\tp
e^{-i\phi Z}\nonumber\\&\equiv&|0\ket\bra0|\tp I+|1\ket\bra1|\tp
e^{2i|\phi|Z}.
\end{eqnarray}
Note that, if necessary, we can obtain a positive exponent in the last
line by conjugating by $I\tp X$.  We introduce the following notation
for a controlled rotation about an arbitrary axis \n,
\begin{equation}
U_\n\equiv|0\ket\bra0|\tp I+|1\ket\bra1|\tp e^{i\n\cdot(X,Y,Z)}.
\end{equation}
In particular, the controlled rotation~(\ref{eq:controlled_rot}) above
is denoted $U_{(0,0,2|\phi|)}$. Conjugation by one-qubit gates on the
second qubit changes the \emph{axis} of rotation but not the
\emph{angle} of rotation: Given $\mathbf{n'}$ such that
$|\n|=|\mathbf{n'}|$, we can find a one-qubit gate $A$ such that
$(I\tp A)U_\n(I\tp A^\dag)=U_\mathbf{n'}$.  A product of two rotations
$U_\n$ and $U_\mathbf{n'}$ is clearly another controlled rotation
$U_\m$.  Both the direction of \m\ and its magnitude vary depending on
\n\ and $\mathbf{n'}$.

In order to implement a \cnot, we need to use $U_{(0,0,2|\phi|)}$
to obtain a total rotation $U_{(0,0,\pi/2)}\equiv\cnot$.  The
first step is to use $U_{(0,0,2|\phi|)}$ a number of times
$q=\l\lfloor\frac{\pi/2}{2|\phi|}\r\rfloor$.  If $\pi/2$ is an
exact multiple of $2|\phi|$, then we are done.  Otherwise, we must
generate a gate to make up the difference; \ie we need to obtain
$U_\m$ with $0<|\m|=\frac{\pi}{2}-2q|\phi|<2|\phi|$. To do this,
we note that we can easily obtain the following controlled
rotations: the zero rotation,
$U_{(0,0,0)}=U_{(0,0,2|\phi|)}U_{(0,0,-2|\phi|)}$, and
$U_{(0,0,4|\phi|)}=U_{(0,0,2|\phi|)}U_{(0,0,2|\phi|)}$.  Choose
\n\ such that $|\n|=2|\phi|$ in which case $U_\n$ is equivalent to
$U_{(0,0,2|\phi|)}$.  The product $U_{(0,0,2|\phi|)}U_\n$ gives
another controlled rotation $U_\m$.  $|\m|$ varies continuously as
a function of \n\ and so, by the Intermediate Value Theorem, it
must pass through all the angles between 0 and $4|\phi|$. As a
consequence it is possible to choose \n\ such that
$|\m|=\frac{\pi}{2}-2q|\phi|$. For any given angle $\phi$, \n\ can
be calculated numerically as the solution to a small set of
equations (these can be found in exercise 4.15
in~\cite{Nielsen00}, see also~\cite{Preskill98c})
\cite{endnote17}.
Therefore, since $U_{(0,0,|\m|)}\equiv U_\m$, the final sequence
is
\begin{equation}
U_{(0,0,\pi/2)}=U_{(0,0,2|\phi|)}^q(I\tp A)U_\m(I\tp A^\dag),
\end{equation}
where $A$ is an appropriate one-qubit gate.

This completes our proof, since it demonstrates that the
imprimitive gate $U$ together with one-qubit gates can be used to
implement a \cnot, which, in turn, can be used to perform
universal quantum computation
\cite{endnote18}.~$\blacksquare$

It is easy to explore some examples of our procedure
using~\cite{Dawson02}.  As an example, suppose we had a gate whose
canonical decomposition yielded $U=e^{i\frac{\pi}{6}Z\tp Z}$. Then
$A_1UA_2UA_3=\cnot$ where the gates $A_j$ are \emph{primitive}:
\begin{eqnarray}
A_1&=&\[\begin{smallmatrix}1&0\\0&-i\end{smallmatrix}\]\tp\(e^{-i\gamma
B}e^{i\beta Y}\),\ A_2=I\tp\(e^{-i\frac{\pi}{6}Z}e^{-i\beta
Y}\),\nonumber\\ A_3&=&I\tp\(e^{-i\frac{\pi}{6}Z}e^{i\gamma B}\),\
B=\({\textstyle\sqrt\frac{3}{5}}Z-{\textstyle\sqrt\frac{2}{5}Y}\),
\end{eqnarray}
where $\beta=\frac{1}{2}\cos^{-1}\frac{1}{3}$ and
$\gamma=\frac{1}{2}\cos^{-1}\frac{1}{\sqrt6}$.

We conclude with a discussion of the optimality of the scheme for
universal quantum computation described in our proof.  We need to
answer two questions: What is the ``optimal'' use of a given gate $U$?
How optimal is our scheme?  We define a scheme to be \emph{optimal} if
it uses $U$ the minimal number of times required to implement a \cnot,
with arbitrary one-qubit gates.  We will see that, although our scheme
is slightly non-optimal in usage of the two-qubit interaction, the
number of one-qubit gates used by our scheme is many orders of
magnitude smaller than the number required by the Hamiltonian
simulation schemes described at the beginning of this Letter.

It follows from~\cite{Hammerer02} (section D1) that the number of
uses of $U$ required to implement the \cnot\ in \emph{any}
protocol using only $U$ and one-qubit gates is bounded below by
$\frac{\pi}{4\theta_\maxi}$ where
$\theta_\maxi=\max\{|\theta_x|,|\theta_y|,|\theta_z|\}$.

To compare with our scheme, we obtain an estimate of the number of
uses of $U$ required to implement a \cnot\ using our scheme.  Recall
that we use a controlled rotation $U_{(0,0,2|\phi|)}$
$q=\l\lfloor\frac{\pi}{8\theta_\maxi}\r\rfloor$ times to implement a
\cnot.  (Recall that we take $\theta_z=\theta_\maxi$.)  Each
controlled rotation uses $U$ twice in general (the special cases
follow along similar lines with small changes in the number of uses of
$U$), and the corrections at the end can require up to four uses of
$U$.  Thus the \cnot\ uses $U$ $2q+4$ times.  The ratio of the number
of uses of $U$ required by our scheme to the minimum possible number
is therefore less than $1+16\theta_\maxi/\pi$, which is between 1 (for
small $\theta_\maxi$) and 5 (for large $\theta_\maxi$).

Returning to the comparison of our result with those on optimal
simulation of Hamiltonians \cite{Bennett01,Vidal01}, note that our
fixed given gate $U$ can be thought of as a fixed given
Hamiltonian which always evolves for the same amount of time
between applications of one-qubit gates.  Although our procedure
is slightly non-optimal in the number of uses of $U$ for large
$\theta_\maxi$, the pay-off in terms of error-control is enormous.
In general, we require only approximately $6q$ one-qubit gates,
and $q$ depends only on the gate $U$, not on the desired accuracy.
In the example given above, only 4 one-qubit gates are required,
compared to the unbounded number required to achieve arbitrary
accuracy in the Hamiltonian simulation procedures.

We have given a simple algorithm \cite{Dawson02} which provides a
near-optimal way of using an \emph{arbitrary} two-qubit entangling
interaction to do universal quantum computation.  Our scheme makes
relatively undemanding requirements on local control, and thus is
likely to be experimentally practical.  Our scheme inverts the
usual challenge facing the designer of a quantum computer: Instead
of having to do delicate, system-specific theoretical calculations
to engineer systems to perform gates such as the \cnot, it will
now be possible for physicists to experimentally determine the
character of the available interaction and then apply our
algorithm to use that interaction to do universal quantum
computation.

\begin{acknowledgements}
We thank Tamyka Bell, Carl Caves, Tim Ralph, and R\"udiger Schack
for helpful comments and suggestions.  A.W.H thanks the Centre for
Quantum Computer Technology at the University of Queensland for
its hospitality and acknowledges support from Army Research
Office. A.G. was supported by The New Zealand Foundation for
Research, Science and Technology.
\end{acknowledgements}



\begin{thebibliography}{15}
\bibitem{endnote17}{The
equations are:
$\cos(|\m|)=\cos^2(2|\phi|)-\sin^2(2|\phi|)\hat{\z}\cdot\hat{\n}$
and
$\sin(|\m|)\hat{\m}=\sin(2|\phi|)\cos(2|\phi|)(\hat{\z}+\hat{\n})-\sin^2(2|\phi|)\hat{\n}\times\hat{\z}$.
These are easily solved using standard numerical techniques.}
\bibitem{endnote18}{Note that our construction could be easily modified in order to simulate any desired two-qubit gate.}
\end{thebibliography}
\end{document}